\documentclass[a4paper,12pt]{elsarticle}

\usepackage[ruled,vlined]{algorithm2e}
\usepackage{url}
\usepackage{hyperref}
\usepackage{array}
\usepackage{multirow}
\usepackage{color}

\usepackage{graphics}
\usepackage{latexsym}
\usepackage{rotating}
\usepackage[utf8]{inputenc}

\usepackage{amsmath, amssymb}
\usepackage{index}
\usepackage{stmaryrd}
\usepackage{leftidx}
\usepackage{mathabx}
\usepackage{tikz}
\usepackage{longtable}
\usetikzlibrary{arrows,matrix,positioning,calc}

\newindex{default}{idx}{ind}{Index}

\newcommand{\noticka}[1]{}
\def\tu#1{\langle #1\rangle}

\newcommand{\up}{{\uparrow}}
\newcommand{\down}{{\downarrow}}

\newcounter{dct}

\newcommand{\blue}[1]{{\color{blue}#1}}
\renewcommand{\blue}[1]{{\color{black}#1}}
\newcommand{\PreserveBackslash}[1]{\let\temp=\\#1\let\\=\temp}
\newcolumntype{C}[1]{>{\PreserveBackslash\centering}p{#1}}

\usepackage{amsthm}

\newtheoremstyle{exampstyle}
  {\topsep} 
  {\topsep} 
  {} 
  {} 
  {\bfseries} 
  {.} 
  {.5em} 
  {} 
\theoremstyle{exampstyle} \newtheorem{example}{Example}

\theoremstyle{remark}
\newtheorem{remark}{Remark}

\begin{document}
\begin{frontmatter}
  \title{LCM from FCA Point of View: A CbO-style Algorithm with Speed-up Features}

\author{Radek Janostik}
\ead{radek.janostik@upol.cz}

\cortext[cor1]{Corresponding author}

\author{Jan Konecny}
\ead{jan.konecny@upol.cz}

\author{Petr Krajča}
\ead{petr.krajca@upol.cz}

\address{Dept.\,Computer Science, Palack\'y University Olomouc \\
	17.\,listopadu 12, CZ--77146 Olomouc \\
	Czech Republic}

\begin{abstract}
LCM is an algorithm for enumeration of frequent closed itemsets in transaction databases.
It is well known that when we ignore the required frequency,
the closed itemsets are exactly intents of formal concepts in Formal Concept Analysis (FCA).
We describe LCM in terms of FCA and show
that LCM is basically the Close-by-One algorithm with multiple speed-up features
for processing sparse data.
We analyze the speed-up features and compare them with those of similar FCA algorithms, 
like FCbO and algorithms from the In-Close family.
\begin{keyword}
algorithm;
formal concept analysis;
frequent closed itemset;
Close-by-One
\end{keyword}
\end{abstract}
\end{frontmatter}

\section{Introduction}
\label{sec:intro}

Frequent closed itemsets in transaction databases are exactly intents in
Formal Concept Analysis (FCA) with sufficient support---cardinality of the corresponding extents. 
If the minimum required support is zero (i.e. any attribute set is considered frequent), one can easily unify these two notions.
LCM ({\bf L}inear time {\bf C}losed itemset {\bf M}iner) is an algorithm for the enumeration of frequent closed itemsets developed by Takeaki Uno \cite{lcm1b,lcm1a,lcm2,lcm3} in 2003--2005. 
It is considered to be one of the most efficient algorithms for this task.\footnote{Its implementations with source codes are available at URL
{\url{http://research.nii.ac.jp/~uno/codes.htm}}.}

We have thoroughly studied Uno's papers and source codes and,
in the present paper, we deliver a complete description of LCM
from the point of view of FCA. Despite the source codes being among
the main sources for this study, we stay at a very comprehensible level
in our description and avoid delving into implementation details.
We explain that the basis of LCM is Kuznetsov's Close-by-One (CbO) \cite{kuznetsov1993fast}.\footnote{Although LCM was most likely developed independently.}
We describe its additional speed-up features and compare them
with those of state-of-art CbO-based algorithms, like FCbO \cite{Outrata2012114}
and In-Close2+  \cite{Andrews:2011:IHP:2032828.2032834,Andrews2015633,Andrews2017,Andrews2018new}.%
\footnote{In the rest of this paper, whenever we write `CbO-based algorithms' we mean CbO, FCbO and In-Close family of algorithms. By version number 2+, we mean the version 2 and higher.}

\begin{remark}[Some subjective notes on our motivations]\
Besides the obvious importance of LCM for FCA\footnote{However, the closed sets are also relevant in other disciplines, like
association rule mining  \cite{agrawal1996fast,bayardo1998efficiently}, condensed representation
of inductive queries \cite{mannila1996multiple}, or logical analysis of data \cite{alexe2007logical,alexe2006spanned}.}, there
are two motivational points for this work.
We separated them into this remark as they are based on our subjective impressions and views.
We ask the reader to excuse the rather subjective tone in this remark.
\begin{itemize}
\item[(a)]
    A significant part of the FCA community is aware of LCM
    and uses Uno's implementation to enumerate formal concepts
    for further processing or for comparison with their
    methods.    
    However, it is our impression that, more often than not,
    the implementation is used merely as a black box.
    We believe that this part of the FCA community would appreciate
    `unwrapping the black box'.
\item[(b)]
    Uno's papers provide quite confusing descriptions of LCM
    and the source codes of the implementations are not written to
    allow easy analysis. Moreover, Uno's implementation and
    description of LCM have some differences; there
    is even an important feature present in the implementation
    which is not described in the papers. We believe
    that a clearer and more complete description would be fruitful.
\end{itemize}
\end{remark}

The paper is structured as follows. 
First, we recall the basic notions and notations of FCA (Section \ref{sec:fca}) we use in the rest of this paper. Then we describe CbO (Section~\ref{sec:cbo}) as it is a basis for description of LCM.
Afterwards, we provide a description of LCM's features (Section~\ref{sec:lcm}), namely,
initial preprocessing of data (Section~\ref{sec:init}),
handling data using arraylists computing all attribute extents at once (Section \ref{sec:arraylists}),
conditional datasets (Section~\ref{sec:conddat}), 
and pruning (Section~\ref{sec:pruning}).
In all of the above, we ignore the requirement of frequency and we describe it separately (Section \ref{sec:freq}). 
\blue{
We provide a tutorial-like description of complete FP-trees used for the denser parts
of data in LCM3 (Section~\ref{sec:fp}).}
Finally, we summarize our conclusions (Section~\ref{sec:conc}).

\medskip

\blue{This paper is an extended version of the conference paper \cite{cla} presented at CLA 2020.}

\section{Formal Concept Analysis}
\label{sec:fca}

\begin{figure}

  \begin{center}
    \includegraphics{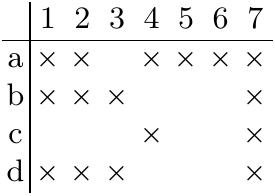}
  \end{center}
  \caption{Formal context with objects a,b,c,d and attributes $1,2,\dots,7$
\label{fig:fc}}
\end{figure}

An input to FCA is a triplet $\tu{X,Y,I}$,
called a \emph{formal context}, where $X$ and $Y$ are non-empty sets of objects and attributes, respectively, and $I$ is a binary relation between $X$ and $Y$; $\tu{x,y}\in I$ means that the object $x$ has the attribute $y$.
 Finite formal contexts are usually depicted as tables, in which rows represent objects, columns represent attributes, and each entry contains a cross if the corresponding object has the corresponding attribute, and is left blank otherwise; see Fig.\,\ref{fig:fc} for an example. 
In this paper, we consider only finite formal contexts.
 
The formal context induces concept-forming operators:
\begin{itemize}
\item[]
${}^\up: \mathbf{2}^X \to \mathbf{2}^Y$ assigns to a set $A$ of objects the set $A^\up$ of all attributes shared by all the objects in $A$.
\item[]
${}^\down:\mathbf{2}^Y \to \mathbf{2}^X$ assigns to a set $B$ of attributes the set $B^\down$ of all objects which share all the attributes in $B$.
\end{itemize}

Formally, 
\begin{align*}
A^\up   &= \{ y \in Y \mid \forall x \in A: \tu{x,y} \in I \}, \quad\text{and}\quad 
B^\down &= \{ x \in X \mid \forall y \in B: \tu{x,y} \in I \}. 
\end{align*}
For singletons, we use shortened notation and write $x^\up$, $y^\down$ instead of $\{x\}^\up$, $\{y\}^\down$, respectively.

In this paper, we assume a set of attributes $Y=\{1,\dots,n\}$.
Whenever we write about lower and higher attributes, we
refer to the natural ordering $\le$ of the numbers in $Y$.

\medskip

A {\em formal concept} is a pair $\tu{A,B}$ of sets $A \subseteq X, B \subseteq Y$, such that $A^\up = B$ and $B^\down = A$.
The first component of a formal concept is called the extent, whilst the second one is called the intent.

\medskip

The compositions ${}^{\up\down}$ and ${}^{\down\up}$ of concept-forming operators are 
closure operators on $2^X$ and $2^Y$, respectively. That is, the composition ${}^{\down\up}$ satisfies
\begin{align}
\text{(extensivity)} && B &\subseteq {B}^{\down\up} \\
\text{(monotony)}    && B \subseteq D &\text{ implies } {B}^{\down\up} \subseteq {D}^{\down\up}\\
\text{(idempotency)} && B^{\down\up} &= B^{\down\up\down\up} 
\end{align}
for all $B,D \subseteq Y$ (analogously for the composition ${}^{\up\down}$).
\bigskip

Sets of attributes satisfying $B = B^{\down\up}$ are called closed sets and they are exactly the intents of formal concepts.



\section{Close-by-One}
\label{sec:cbo}

In the context of FCA, the foundation of LCM is CbO.\footnote{It is called a backtracking algorithm with prefix-preserving closure extensions in Uno's papers.}
Therefore, we firstly turn our attention to CbO.

\bigskip

We start the description of CbO with a na\"ive algorithm for generating all closed sets.
The na\"ive algorithm
traverses the space of all subsets of $Y$, 
each subset is checked for closedness and is outputted.
This approach is quite inefficient as the number of closed subsets is typically significantly smaller than the number of all subsets.

\begin{algorithm}
\DontPrintSemicolon\LinesNumbered
\SetKwInOut{Input}{input}\SetKwInOut{Output}{output}
\SetKwProg{Fn}{def}{\string:}{}
\SetKwFunction{FRecurs}{GenerateFrom}%
\Fn{\FRecurs{$B$, $y$}}{
\Input{$B$ -- set of attributes\\ $y$ -- last added attribute}
\If{$B = B^{\down\up}$}{%
  {\bf print}({$B$})
}
\For{$i \in \{y+1,\dots,n\}$}{
$D \gets B \cup \{i\}$\;
\FRecurs{$D$, $i$}\;
}
{\bf return}
}
\FRecurs{$\emptyset$, $0$}
\caption{Na\"\i{}ve algorithm to enumerate closed subsets\label{alg:subsets}}
\end{algorithm}

The algorithm is given by a recursive procedure \texttt{GenerateFrom}, which accepts two attributes: 
\begin{itemize}
\item[$\bullet$] $B$ -- the set of attributes, from which new sets will be generated.
\item[$\bullet$] $y$ -- the auxiliary argument to remember the highest attribute in $B$.
\end{itemize}
The procedure first checks the input set $B$ for closedness and prints it if it is closed (lines 1,2). Then, for each
attribute $i$ higher than $y$:
\begin{itemize}
\item[$\bullet$]  a new set is generated by adding the attribute $i$ into the set $B$ (line~4);
\item[$\bullet$]  the procedure recursively calls itself to process the new set (line~5).
\end{itemize}
The procedure is initially called with an empty set and zero as its arguments.

\medskip

\begin{figure}
\begin{center}
\includegraphics{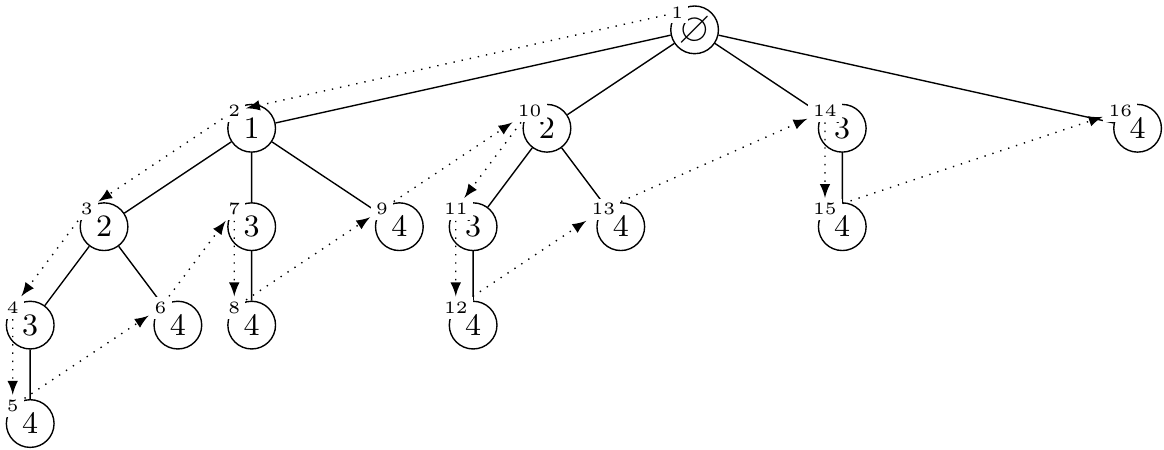}
\end{center}
\caption{Tree of all subsets of $\{1,2,3,4\}$. Each node represents a unique set containing all elements in the path from the node to the root. 
The dotted arrows and small numbers represent the order of traversal performed by the algorithm for generating all subsets.
\label{fig:tree}}
\end{figure}

The na\"ive algorithm represents a depth-first sweep through the tree of all subsets of $Y$ (see Fig.\,\ref{fig:tree}) and printing the closed ones.

In the tree of all subsets (Fig.\,\ref{fig:tree}), 
each node is a superset of its predecessors.
We can use the closure operator ${}^{\down\up}$ to skip non-closed sets. In other words, to make jumps in the tree to closed sets only.
Instead of simply adding an element to generate a new subset
$D \gets
B \cup \{i\}$,
CbO adds the element and then closes the set
\begin{equation}
\label{eq:D}
D \gets
(B \cup \{i\})^{\down\up}.
\end{equation}
We need to distinguish the two outcomes of the closure \eqref{eq:D}.
Either 
\begin{itemize}
\item
the closure contains some attributes lower than $i$ which are not included in $B$, i.e.
$D_i \neq B_i$ 
where
$D_i = D \cap \{1, \dots, i-1\},\,B_i = B \cap \{1, \dots, i-1\}$;
\item
or it does not, and we have
$D_i = B_i$.
\end{itemize}
The jumps with $D_i \neq B_i$ are not desirable because they land on
a closed set which was already processed or will be processed later (depending on the direction of the sweep). 
CbO does not perform 
such jumps.
The check of the condition $D_i = B_i$ is called a {\em canonicity test}.



%
%

%
%
%
%
%
%
%
%

\RestyleAlgo{ruled}
\begin{algorithm}
\DontPrintSemicolon\LinesNumbered
\SetKwInOut{Input}{input}\SetKwInOut{Output}{output}
\SetKwProg{Fn}{def}{\string:}{}
\SetKwFunction{FRecurs}{GenerateFrom}%

\Fn{\FRecurs{$A$, $B$, $y$}}{
\Input{$A$ -- extent\\$B$ -- set of attributes\\ $y$ -- last added attribute}


$D \gets A^\up$\;

\If{$D_y \neq B_y$}{%
{\bf return}
}

{\bf print}({$\tu{A,D}$})\;

\For{$i \in \{y+1,\dots, n \} \,\setminus\, D$}{
$C \gets A \cap {i}^\down$\;
\FRecurs{$C$, $D \cup \{i\}$, $i$}\;
}
{\bf return}
}

\FRecurs{$X$, $\emptyset$, $0$}

\caption{Close-by-One\label{alg:cbo}}
\end{algorithm}

One can see the pseudocode of CbO in Algorithm~\ref{alg:cbo}.
In addition, we pass an extent $A$ to the procedure \texttt{GenerateFrom}
and the closed set $D (= B^{\down\up})$ is computed as $A^\up$ (line 1), which is more efficient.
Then the canonicity test is performed. If it fails, the procedure ends (lines 2, 3). Otherwise, we print the  concept and continue with the generation of its subtree.

\begin{remark}
What we show in Algorithm~\ref{alg:cbo} is not the usual pseudocode 
of CbO presented in literature. 
We made some superficial changes to emphasize the link between
CbO and LCM, which will be apparent later.
Specifically: 
\begin{itemize}
\item[(a)]
The closure and the canonicity test (lines 1--3)
are usually performed before the recursive invocation (line 7), not as the first
steps of the procedures. 
\item[(b)]
The main loop (line 5) of CbO usually processes the attributes in the ascending order, which corresponds to a left-to-right depth-first sweep through the tree of all subsets
(Fig.\,\ref{fig:tree}). In actual fact, for CbO there is no reason for a particular order 
of processing attributes. 
\end{itemize}
\end{remark}

%
%


\section{Features of LCM}
\label{sec:lcm}

There are three versions of the LCM algorithm:
\begin{description}
\item[LCM1] is CbO with arraylist representation of data and computing of all extents at once (described in Section~\ref{sec:arraylists}), 
data preprocessing (described in Section~\ref{sec:init}),
and
using of diffsets \cite{zaki2003fast} to represent extents for dense data (this is not present in later versions).
\item[LCM2] is LCM1 (without diffsets) with conditional databases (described in Section~\ref{sec:conddat})
\item[LCM3] is LCM2 which uses a hybrid data structure to represent a context. The data structure
uses a combination of FP-trees and bitarrays, called a complete \blue{FP-tree (described in Section~\ref{sec:fp})}, to handle the most dense attributes.
Arraylists are used for the rest, the same way as in the previous versions.
\end{description}

\noindent
In this paper, we describe all features present in LCM2 \blue{and LCM3}.

\subsection{Initialization}
\label{sec:init}
To speed the computation up, LCM initializes the input data as follows:
\begin{itemize}
\item[$\bullet$]
removes empty rows and columns,
\item[$\bullet$]
merges identical rows,
\item[$\bullet$]
sorts attributes by cardinality ($|y^\down|$) in the descending order,
\item[$\bullet$] 
sorts objects by cardinality $(|x^\up|)$ in the descending order.
\end{itemize}


In the pseudocode in Algorithm~\ref{alg:lcm2} 
(later in the paper),
the initialization is not shown and it is supposed that it is run before the first invocation
of the procedure \texttt{GenerateFrom}.

\paragraph{FCA aspect:}{

The attribute sorting is well known to  most likely cause a smaller number of computations of closures in CbO-based algorithms \cite{krajca2010advances,Andrews:2011:IHP:2032828.2032834,Andrews2015633}. This feature is included in publicly available implementations of In-Close4 and FCbO.

The object sorting is a different story. Andrews  \cite{Andrews:2011:IHP:2032828.2032834} tested 
the performance of In-Close2
and concluded that 
lexicographic order tends to significantly reduce L1 data cache misses. 
However, the test were made for bitarray representation of contexts.
 
The reason for object sorting in LCM is probably that a lesser amount of 
inverses occurs in a computation of a union of rows (shown later \eqref{eq:Nunion}), which is consequently 
easier to sort.
Our testing with Uno's implementation of LCM 
did not show any difference in runtime for unsorted and sorted objects
when attributes are sorted. In the implementation of LCM3, the object sorting is not present.


\begin{remark}
In examples in the present paper, we do not use sorted data, in order to keep the examples small.
\end{remark}

\subsection{Ordered arraylists and occurrence deliver}
\label{sec:arraylists}

LCM uses arraylists\footnote{Whenever we write arraylist, we mean ordered arraylists.} 
as data representation of
the rows of the context. It is directly bound
to one of the LCM's main features -- {\em occurrence deliver}:

LCM computes extents 
$A \cap i^\down$ (line 6 in Algorithm~\ref{alg:cbo})
all at once using a single traversal through the data.
Specifically, it sequentially traverses through all rows $x^\up$
of the context and whenever it encounters an attribute~$i$,
it adds $x$ to an initially empty arraylist -- {\em bucket} -- for $i$ (see Fig.\,\ref{fig:occdel}).
As LCM works with conditional datasets (see Section~\ref{sec:conddat}),
attribute extents correspond to extents $A \cap i^\down$ (see Algorithm~\ref{alg:cbo}, line 6). This is also known as {\em vertical format} in DM algorithms; the buckets are also known as {\em tidlists}.

\begin{figure}
\includegraphics{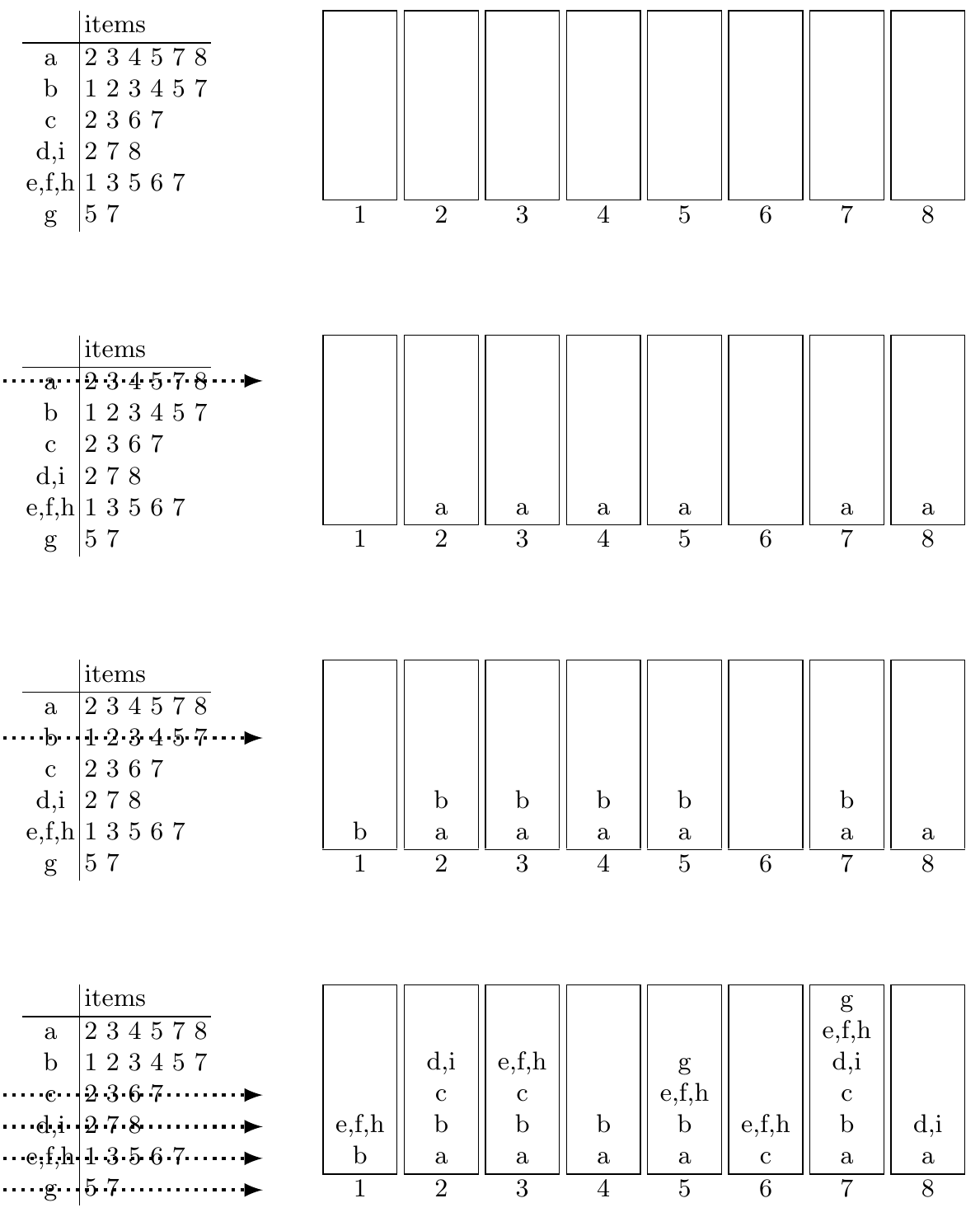}
\caption{Occurrence deliver in LCM. \label{fig:occdel}}
\end{figure}

\bigskip

LCM generates childs of each node from right to left.
That way, it can reuse the memory for extents (buckets).
For example, when computing extents in the node $\{2\}$, 
that is $\{2,3\}^\down$ and $\{2,4\}^\down$, the algorithm
can reuse the memory used by extents $\{3\}^\down$ and $\{4\}^\down$,
because $\{3\}$ and $\{4\}$ (and their subtrees) are already finalized (see Fig.\,\ref{fig:RFS}).

\begin{figure}
\begin{center}
\includegraphics{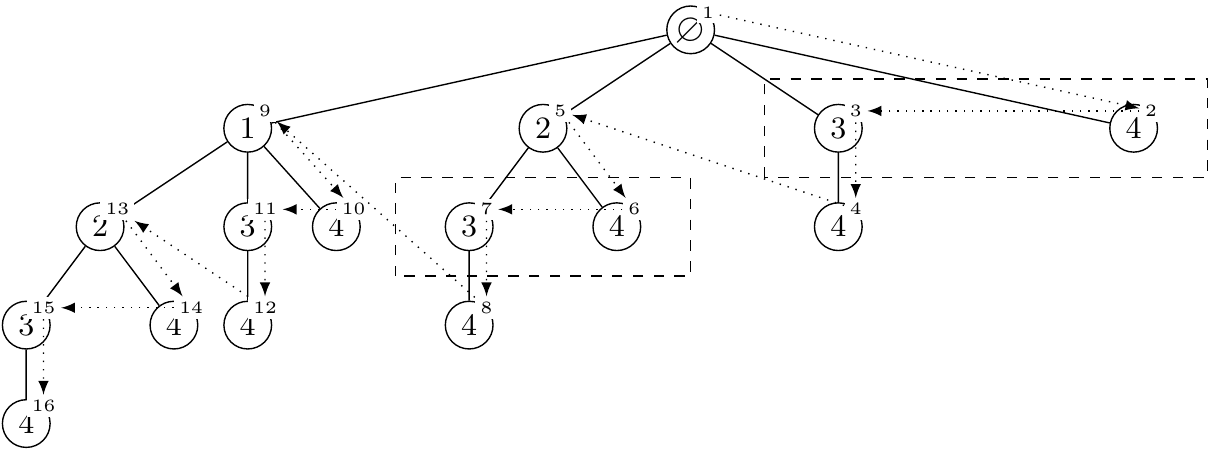}
\end{center}
\caption{Demonstration of bucket reuse in LCM with right-first sweep.\label{fig:RFS}}
\end{figure}

%
%

\begin{figure}
\begin{center}


  \includegraphics{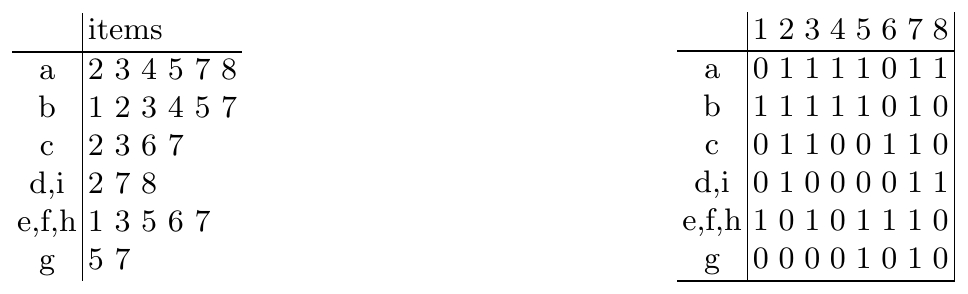}
\end{center}
\caption{Data representation for contexts: arraylists (left), bitarrays (right).}
\end{figure}

\paragraph{FCA aspect:}
In FCA, the CbO-based algorithms do not specify data representation used for 
handling contexts, sets of objects, and sets of attributes.
This is mostly considered a matter of specific implementations (see Remark~\ref{rem:datastruct}).
Generally, the data representation issues are almost neglected in literature
on FCA. The well-known comparison study \cite{serser} of FCA algorithms mentioned the
need to study the influence of data structures on practical performances of FCA
algorithms 
but it does not pay attention to that particular issue.
The comparison study \cite{krajca2009comparison} provided the first steps to an answer for this need.\footnote{Paper \cite{krajca2009comparison} compares bitarrays, sorted linked lists, arraylists, binary search trees, and hash tables.}
The latter paper concludes that binary search trees or linked lists are good choices for large or sparse datasets,
while bitarray is an appropriate structure for small or dense datasets. Arraylists did not perform
particularly  
well in any setting. However, this comparison did not assume other features 
helpful for this data representation, like conditional databases (see Section~\ref{sec:conddat}) and 
computation of all required attribute extents in one sweep by occurrence deliver.
More importantly, the minimal tested density is 5\,\%, which is still very dense
in the context of transactional data.


%
%

\begin{remark}
\label{rem:datastruct}
Available implementations of 
FCbO\footnote{Available at \url{http://fcalgs.sourceforge.net/}.} and In-Close\footnote{Available at \url{https://sourceforge.net/projects/inclose/}.}
utilize bitarrays for rows of contexts, and sets of attributes, and arraylists for sets of objects.
\end{remark}

\subsection{Conditional Database and Interior Intersections}
\label{sec:conddat}

LCM reduces the database for the recursive invocations
of \texttt{GenerateFrom} to so-called conditional databases.

Let
${\cal K} = \tu{X,Y,I}$ be a formal context, 
$B$ be an intent, and $y$ be the attribute used to build $B$.
The conditional database (context) ${\cal K}_{B,y}$ w.r.t. $\tu{B,y}$ 
is created from ${\cal K}$ as follows:
\begin{itemize}
\item[(a)]
First, remove objects from ${\cal K}$ which are not in the corresponding extent $A = B^\down$.
\item[(b)]
Remove attributes which are full or empty.
\item[(c)]
Remove attributes lesser than $y$. 
\footnote{In the implementation, when the database is already too small (less than 6 objects, and less than 2 attributes), steps (c)--(d) are skipped.}
\item[(d)]
Merge identical objects together.
\item[(e)] 
Put back attributes removed in step (c);
set each new object created in step (d) by merging objects $x_1,\dots,x_k \in X$ to have attributes common to the objects $x_1,\dots,x_k$.
That is, the merged objects are intersections of the original objects.
The part of the context added in this step is called an interior intersection.
\end{itemize}
For an example, see Fig.\,\ref{fig:conditionalDB}.

\begin{figure}
  \begin{center}
    \includegraphics{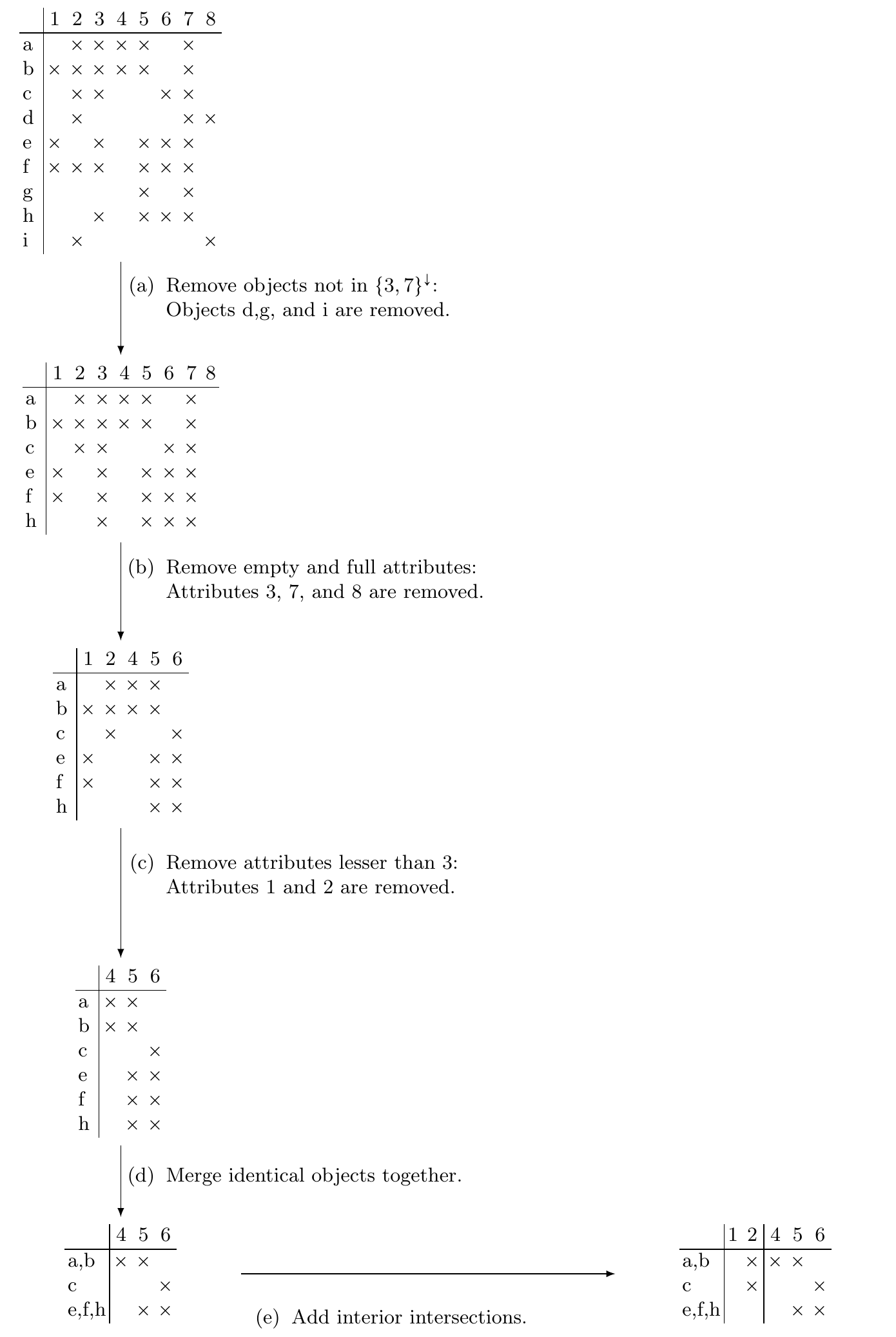}
\end{center}
\caption{Obtaining contitional context ${\cal K}_{B,y}$ for attribute set $B = \{3,7\}$ and attribute $y=3$.\label{fig:conditionalDB}}
\end{figure}

Alternatively, we can describe conditional databases with interior intersections as:
\begin{itemize}
\item[$\bullet$]
 Restricting the context ${\cal K}$ to objects in $A$ and attributes in $N$ where

\begin{equation}
\label{eq:Nunion}
N=\left(\bigcup_{x \in A} x^\up 
\right)\,\setminus\,A^\up.
\end{equation}

This covers the steps (a)--(c).
\item[$\bullet$]
Subsequent merging/intersecting of those objects which have the same incidences with attributes in $\{1,2,\dots,y-1\}$. This covers the steps (d)--(e).
\end{itemize}

Note, that from the conditional database (with interior intersections) 
${\cal K}_{B,i}$ 
all intents in the subtree of $B$
can be extracted. More specifically, 
if $D$ is an intent of the conditional database and it passes the canonicity test
(tested on interior intersections) then $D \cup B$ is an intent of the original context 
which passes the canonicity test; i.e. is in the subtree of $B$.

\medskip

In pseudocode in Algorithm~\ref{alg:lcm2} 
(later in the paper),
the creation of the conditional databases with interior intersections is represented by the procedure named
$\texttt{CreateConditionalDB}({\cal K},A,N,y)$.

\paragraph{FCA aspect:} CbO-based algorithms do not utilize conditional databases. 
However, we can see partial similarities with features of CbO-based algorithms.

First, all of the algorithms skip attributes in $B$ and work only with part of the formal context 
given by $B^\down$ and $Y\,\setminus\,B$. This corresponds to step (a) and the
first part of step (b) (full attributes).

Second, the removal of empty attributes in step (b) utilizes basically the same idea as in 
In-Close4 \cite{Andrews2017}: if the present extent $A$ and an attribute extent $i^\down$
have no common object, we can skip the attribute $i$ in the present subtree.
In FCbO and In-Close3, such attribute would be skipped due to pruning (see Section~\ref{sec:pruning}).

Steps (c)--(e) have no analogy in CbO algorithms. 

\begin{algorithm}[t]
\DontPrintSemicolon\LinesNumbered
\SetKwInOut{Input}{input}\SetKwInOut{Output}{output}
\SetKwProg{Fn}{def}{\string:}{}
\SetKwFunction{FRecurs}{GenerateFrom}%
\SetKwFunction{OccDel}{OccurenceDeliver}%
\SetKwFunction{makecond}{CreateConditionalDB}%
\SetKwFunction{freccalc}{Frequencies}%

\Fn{\FRecurs{$A$, $B$, $y$, ${\cal K}$}}{
\Input{$A$ -- extent\\$B$ -- set of attributes\\ $y$ -- last added attribute \\ ${\cal K}$ -- input context/conditional database}


\medskip

$N \gets \left(\bigcup_{x\in A} x^{\up}\right) \,\setminus\, B$\;
$\{ n_i \mid i \in N \} \gets $\freccalc{${\cal K}, N$}\;


\For{$i\in N, i < y$}{
\If{$n_i = |A|$}{%
{\bf return}\;
}
}
\For{$i\in N, i > y$}{
\If{$n_i = |A|$}{%
$B \gets B \cup \{ i \}$\;
$N \gets N \,\setminus\, \{ i \}$
}
}
\medskip

{\bf print}({$\tu{A,B}$})\;

\medskip
${\cal K}' \gets$ \makecond{${\cal K}$, $A$, $N$, $y$}\;
$\{ C_i \mid i \in N \} \gets$ \OccDel{${\cal K}'$}\;

\medskip

\For{$i\in N, i > y$, {\em (in descending order)}}{
\FRecurs{$C_{i}$, $B\cup\{i\}$, $i$, ${\cal K}'$}\;
}
{\bf return}\;
}
\medskip

\FRecurs{$X$, $X^{\up}$, $0$, $\tu{X,Y,I}$}

\caption{LCM (without pruning)\label{alg:lcm2}}
\end{algorithm}

\subsection*{Pseudocode of LCM without pruning}

At this moment, we present pseudocode of LCM (Algorithm~\ref{alg:lcm2})
with above-described features.
As in the case for CbO, the algorithm is given by recursive procedure \texttt{GenerateFrom}.
The procedure takes four arguments: an extent $A$, a set of attributes $B$, the last attribute $y$ added to $B$, and a (conditional) database ${\cal K}$).
The procedure performs the following steps:\\[4pt]
\begin{tabular}{rp{10cm}}
(line 1) & The set $N$ \eqref{eq:Nunion} of non-trivial attributes is computed.\\
(line 2) & The frequencies of all attributes in $N$ are computed, this is made by a single traversal through $\cal K$ similar to the occurence deliver (described in Section~\ref{sec:arraylists}). \\
(lines 3--5) & The loop checks whether any attribute in $N$ lesser than $y$ has frequency equal to $|A|$. 
If so, the attribute causes the canonicity test to fail, therefore we end the procedure. \\
(lines 6--9) & The loop closes $B$ (and updates $N$) based on the computed frequencies. \\
(line 10) & As the cannonicity is checked and $B$ is closed, the pair $\tu{A,B}$ is printed out.\\
(line 11) & The conditional database ${\cal K}_{B,y}$ (described in Sec.~\ref{sec:conddat}) is created. \\
(line 12) & Attribute extents from ${\cal K}_{B,y}$ are computed using occurence deliver (described in Section~\ref{sec:arraylists}).\\
%
(lines 13--14) & The procedure \texttt{GenerateFrom} is recursively called for attributes in $N$
with the conditional database ${\cal K}_{B,y}$ and the corresponding attribute extent.
\end{tabular}

\subsection{Bonus Feature: Pruning}
\label{sec:pruning}

The jumps using closures in CbO significantly reduce the number of visited nodes
in comparison with the na\"ive algorithm. The closure, however, becomes the most
time consuming operation in the algorithm. The pruning technique in LCM%
\footnote{Pruning is not described in papers on LCM, however, it is present in the implementation of LCM2.}
avoids the
computation of some closures based on the monotony property:
for any set of attributes $B,D \subseteq Y$ satisfying $B \subseteq D$, we have that
\begin{equation}
\label{eq:pruneproperty}
j \in (B \cup \{i\})^{\down\up}\quad\text{implies}\quad j \in (D \cup \{i\})^{\down\up}.
\end{equation}
When $i,j \notin D$ and $j < i$, the implication \eqref{eq:pruneproperty}
says that if $j$ causes 
$(B \cup \{i\})^{\down\up}$ to fail
the canonicity test,
then it also causes
$(D \cup \{i\})^{\down\up}$ to fail
the canonicity test.
That is, if we store that $(B \cup \{i\})^{\down\up}$ failed, we can use it to skip the
computation of the closure $(D \cup \{i\})^{\down\up}$ for any $D \supset B$ with $j \notin D$.
We demonstrate this in the following example.

\begin{example}
\label{ex:fcbo_pruning}
Consider the following formal-context.
\begin{center}
\includegraphics{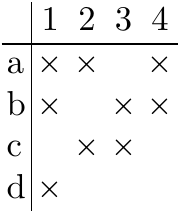}
\end{center}

\noticka{
\begin{center}
\begin{tikzpicture}[xscale=1.5]
\node (0L) at (1,0) {$\emptyset$};
\node[draw,circle] (0) at (0L) {~~~};

\node (2L) at (-2,-1) {$1$};
\node[draw,circle] (2) at (2L) {~~~};

\node (3L) at (0,-1) {$2$};
\node[draw,circle] (3) at (3L) {~~~};

\node (4L) at (2,-1) {$3$};
\node[draw,circle] (4) at (4L) {~~~};

\node (5L) at (4,-1) {$4$};
\node[draw,circle] (5) at (5L) {~~~};

%
%
%
%
%
%
%

\draw (0) -- (2);
\draw (0) -- (3);
\draw (0) -- (4);
\draw (0) -- (5);

%
%
%
%
%

\end{tikzpicture}
\end{center}
}

Consider the tree of all subsets in Fig.\,\ref{fig:tree} (ignoring the left-first sweep order for now).
The rightmost branch of the tree represents adding the attribute $4$ into an empty set.
We can easily see, that 
\begin{align}
\label{eq:414}
\{ 4 \}^{\down\up} = \{1,4\},
\end{align}
and, therefore, the canonicity test $B_i = D_i$ fails. In this case,
we have
$B_i = \emptyset_4 = \emptyset$ 
while
$D_i = \{1,4\}_4 = \{1\}$.

Notice that \eqref{eq:414} gives us information about the actual set (an empty set in this case):
adding attribute $4$
causes that attribute $1$ is in the closed set. 
Due to \eqref{eq:pruneproperty} 
 this holds
true for any superset of the actual set.
This information is then reused for the supersets. Specifically, 
 for sets $\{2\}, \{3\}$, or $\{2,3\}$, adding  attribute $4$ causes
that  attribute 1 is present in the closed set and, consequently, causes 
a failing the canonicity test.
Figure~\ref{fig:cbo_pruning} shows the described situation.

\begin{figure}
 \includegraphics{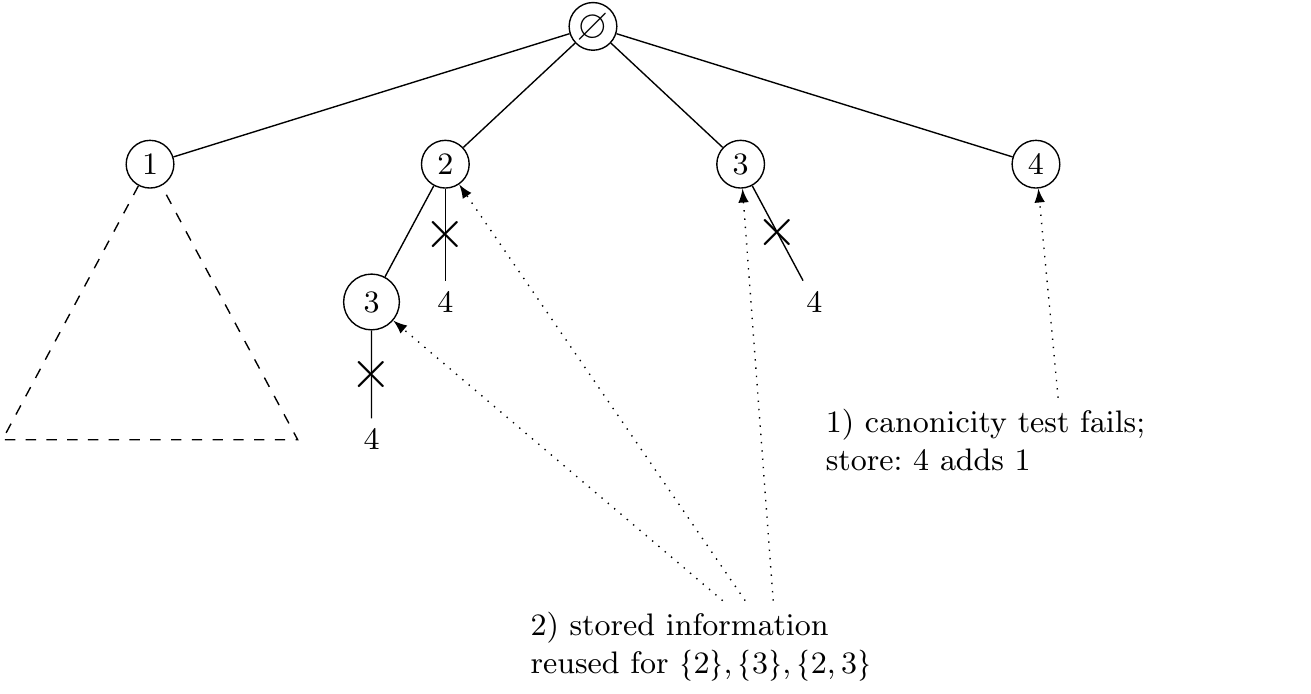}
  \caption{Reuse of failed canonicity test information\label{fig:cbo_pruning}}
\end{figure}
\end{example}

LCM utilizes the above idea in the following way: 
\begin{itemize}
\item[(p0)]
Whenever the canonicity
test fails for $(B \cup \{i\})^{\down\up}$
and $j$ is the smallest attribute in $(B \cup \{i\})^{\down\up} \setminus B$,
we store the rule ``$i$ adds $j$''. In the pseudocode (Algorithm~\ref{alg:lcm:complete}),
this is achieved through the return value of the procedure \texttt{GenerateFrom}.
The procedure returns the lowest attribute which caused
the canonicity to fail (line 6) or 0 if it passed the canonicity test (line 20).
The returned value is used to form a pruning rule, which is to be stored (lines 17,18).
\item[(p1)]
At the beginning of \texttt{GenerateFrom}, i.e. when descending to a subtree,
all rules having the last added attribute (argument $y$) on the right side are removed from the stored rules.
In the pseudocode, this is done by a subroutine called \texttt{RemoveRulesByRightSide} (line 3).
\item[(p2)]
At the end of \texttt{GenerateFrom}, i.e. when backtracking from the current subtree,
all rules from this call are removed.
In the pseudocode, this is done by a subroutine called \texttt{RemoveAllRulesAddedThisCall} (line 19).
\item[(p3)]
Before computing a closure $(D \cup \{i\})^{\down\up}$ in a subtree of $B$,
we check the stored rules to find whether adding $i$ does not add any attribute
which causes the canonicity test to fail. Due to how the rules are handled in the previous items, (p0)--(p2), it is sufficient
to check whether there is any rule having $i$ on the left side.
In the pseudocode, this is done by a subroutine called \texttt{CheckRulesByLeftSide} (line 15).
\end{itemize}

\begin{algorithm}
\DontPrintSemicolon\LinesNumbered
\SetKwInOut{Input}{input}\SetKwInOut{Output}{output}
\SetKwProg{Fn}{def}{\string:}{}
\SetKwFunction{FRecurs}{GenerateFrom}%
\SetKwFunction{OccDel}{OccurenceDeliver}%
\SetKwFunction{makecond}{CreateConditionalDB}%
\SetKwFunction{freccalc}{Frequencies}%
\SetKwFunction{addrule}{AddRule}%
\SetKwFunction{removerulesaddedthiscall}{RemoveAllRulesAddedThisCall}
\SetKwFunction{removerulesrightside}{RemoveRulesByRightSide}
\SetKwFunction{checkrulesleftside}{CheckRulesByLeftSide}

\Fn{\FRecurs{$A$, $B$, $y$, ${\cal K}$}}{
\Input{$A$ -- extent\\$B$ -- set of attributes\\ $y$ -- last added attribute \\ ${\cal K}$ -- conditional database}

\vspace{.25cm}

\medskip

$N \gets \left(\bigcup_{x\in A} x^{\up}\right) \setminus B$\;
$\{ n_i \mid i \in N \} \gets $\freccalc{${\cal K}, N$}\;

\vspace{.25cm}

\removerulesrightside{$y$}\;

\vspace{.25cm}

\For{$i\in N, i < y$}{
\If{$n_i = |A|$}{%
{\bf return} $i$\;
}
}
\For{$i\in N, i > y$}{
\If{$n_i = |A|$}{%
$B \gets B \cup \{ i \}$\;
$N \gets N \setminus \{ i \}$
}
}

\medskip

{\bf print}({$\tu{A,B}$})\;

\medskip
${\cal K}' \gets$ \makecond{${\cal K}$, $A$, $N$, $y$}\;
$\{ C_i \mid i \in N \} \gets$ \OccDel{${\cal K}'$}\;

\medskip

\For{$i\in N, i > y$, {\em (in descending order)}}{
\If{\checkrulesleftside{$i$}}{
$j \gets$ \FRecurs{$C_{i}$, $B\cup\{i\}$, $i$, ${\cal K}'$}\;
\If{$j > 0$}{\addrule(``$i$ adds $j$'')}
}}
\removerulesaddedthiscall{}\;
{\bf return} $0$\;}
\medskip

\FRecurs{$X$, $X^{\up}$, $0$, $\tu{X,Y,I}$}
\caption{LCM\label{alg:lcm:complete}}
\end{algorithm}

\paragraph{FCA aspect:}{%
Similar pruning techniques are also present in FCbO and In-Close3 and higher:

\begin{itemize}
\item
FCbO, In-Close3: stores rules of the form ``$i$ gives set $A$''.
\item
In-Close4: stores rules of the form ``$i$ gives an empty extent''.
\item
In-Close5: stores rules of the form ``$i$ adds an attribute which makes the canonicity test fail'' and rules of In-Close4.
\end{itemize}
All the FCA algorithms utilize only steps (p0), (p2), and (p3); none of them performs (p1).

LCM's pruning is weaker than the pruning in FCbO and In-Close3, yet is stronger than the pruning in In-Close4 and In-Close5.\footnote{In the conference paper, we incorrectly claimed that LCM's pruning is incomparable with the pruning in In-Close5.}:

\begin{center}
In-Close4 $<$ In-Close5 $<$ LCM $<$ FCbO $=$ In-Close3.
%
\end{center}

\section{Frequency Counting}
\label{sec:freq}
In the previous sections, we do not take into account the 
frequency of itemsets, as in FCA, the frequency is not usually
assumed.
However, the implementations of FCbO and In-Close4
allow us to pass minimum support as a parameter, and then
enumerate only frequent intents.

The CbO-based algorithms utilize a simple {\em apriori} principle:
if a set is infrequent then all its supersets are infrequent.
That directly translates into the tree-like computation of the algorithms --
if a node represents an infrequent set, then its subtree does not contain
any frequent set.

In LCM, we make the following modification of the features described in
Section~\ref{sec:lcm}.
\paragraph{Data representation} -- each arraylist is accompanied with a weight,
i.e. number of objects it corresponds to. 

\paragraph{Initialization} -- additionally, infrequent attributes are removed 
and the weights of rows are set to reflect the number of merged rows.
\blue{The weight of unique rows is set to 1}.

\paragraph{Conditional databases} -- in the step (b), infrequent attributes
are removed as well; and in the step (d), the weights of rows are updated.


%
%

\section{\blue{FP-trees in LCM3}}
\label{sec:fp}
Here, we provide a tutorial-like description of the complete FP-trees with inner intersections used in the hybrid data structure of LCM3.

Throughout this section, we assume that the input data does not contain empty rows -- they were removed by the initialization procedure (Section~\ref{sec:init}).

\subsection{Simple FP-tree}

An FP-tree \cite{han2000mining} (also called a prefix tree) is a compressed representation of the input data. It is constructed
by reading the data set one row at a time and mapping each of them
onto a path in the FP-tree. The path contains one node for each attribute of the 
corresponding row, in descending order w.r.t. cardinality of the attributes.
Multiple transactions can share part of their paths. 
Besides their attributes, the nodes contain a number of rows which share the node.
For practicality, nodes corresponding to the same attributes
are linked to lists.

It is not necessary to use the descending order, but it tends to make 
the FP-tree more condense. Note that the attributes are sorted
by their cardinality in ascending order 
in the pre-processing step (Section~\ref{sec:init})
and then processed from right to left. 
This way, they are indeed processed in the descending order.

\medskip

The following examples depict the construction of an FP-tree and extracting a conditional dataset
from it.

\begin{example}[Initial construction of an FP-tree]
\label{ex:aa}
Consider the data depicted in Fig.\,\ref{fig:fp:1}. 
The attributes, sorted by their cardinality, follow the alphabetical order:
\begin{center}
(smallest) {\bf a} $<$ {\bf b} $<$ {\bf c} $<$ {\bf d} $<$ {\bf e} (largest).
\end{center}
Therefore, for instance, the first row will be added into the FP-tree as path (root)--({\bf e})--({\bf d}).
Figure~\ref{fig:fp:first3} shows the FP-tree after adding the first, second, and third row.
Figure~\ref{fig:fp:1}\,(right) shows the FP-tree after adding all the rows.

\begin{figure}
\begin{center}
   \includegraphics{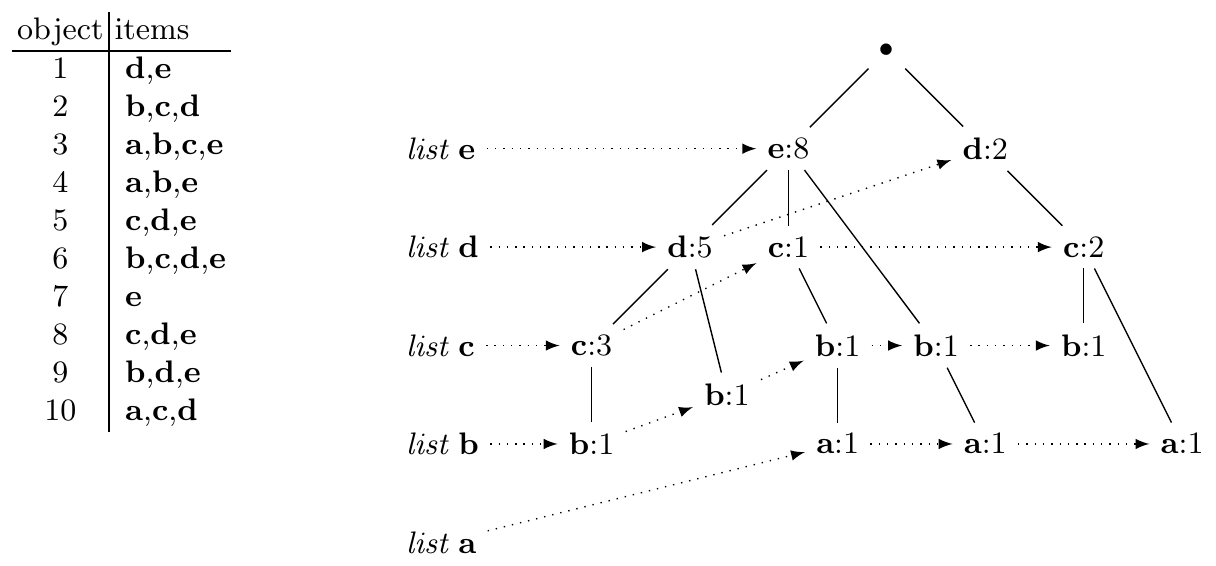}
\end{center}
\caption{Data (left) and its FP-tree representation (right).\label{fig:fp:1}}
\end{figure}
\end{example}

\begin{figure}
\begin{center}
\includegraphics{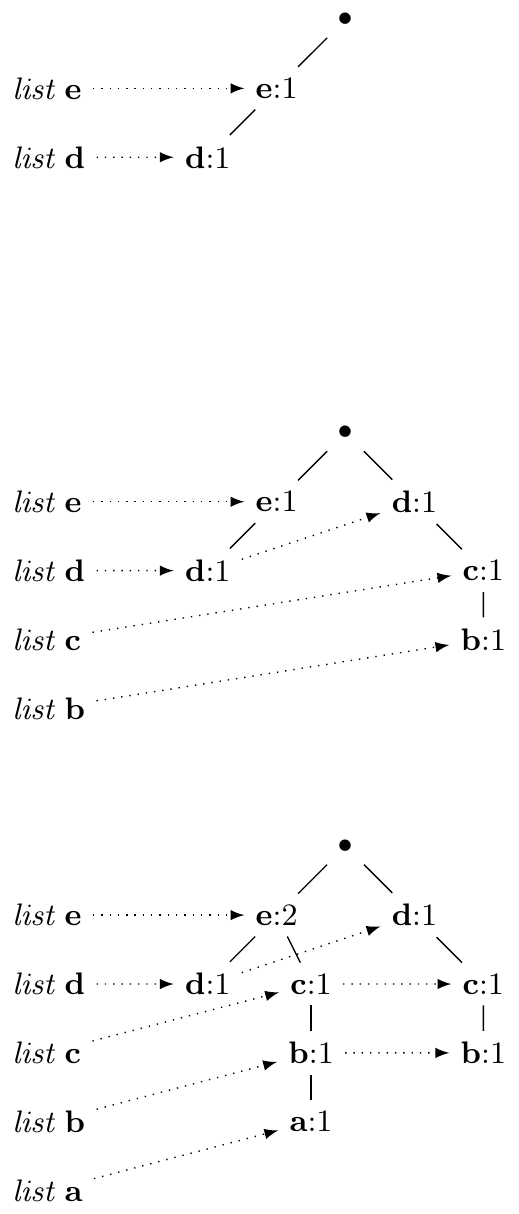}
\end{center}
\caption{First three steps of the construction of the FP-tree for data from Fig.\,\ref{fig:fp:1}.\label{fig:fp:first3}}
\end{figure}



\begin{example}[Extraction of conditional FP-tree]
\label{ex:condi}
Let us construct a conditional FP-tree for attribute {\bf a} from the FP-tree in Fig.\,\ref{fig:fp:1} as follows:
\begin{itemize}
\item[(a)] Take only the paths which contain the attribute {\bf a}. 
\item[(b)] Update numbers to match the numbers of {\bf a}. 
\item[(c)] Remove nodes {\bf a} (and their successors, which is trivial in this case).
\item[(d)] Remove infrequent items, if frequency is considered. In Fig.\,\ref{fig:fp:conda},
			  `{\it list} {\bf d}' contains only one node with count 1. That means that only 
			  one row contains $\{\text{\bf a},\text{\bf d}\}$. If this is seen as infrequent, 
			  we can remove the {\bf d}'s node.
\end{itemize}
\end{example}

\begin{figure}
\begin{center}
\includegraphics{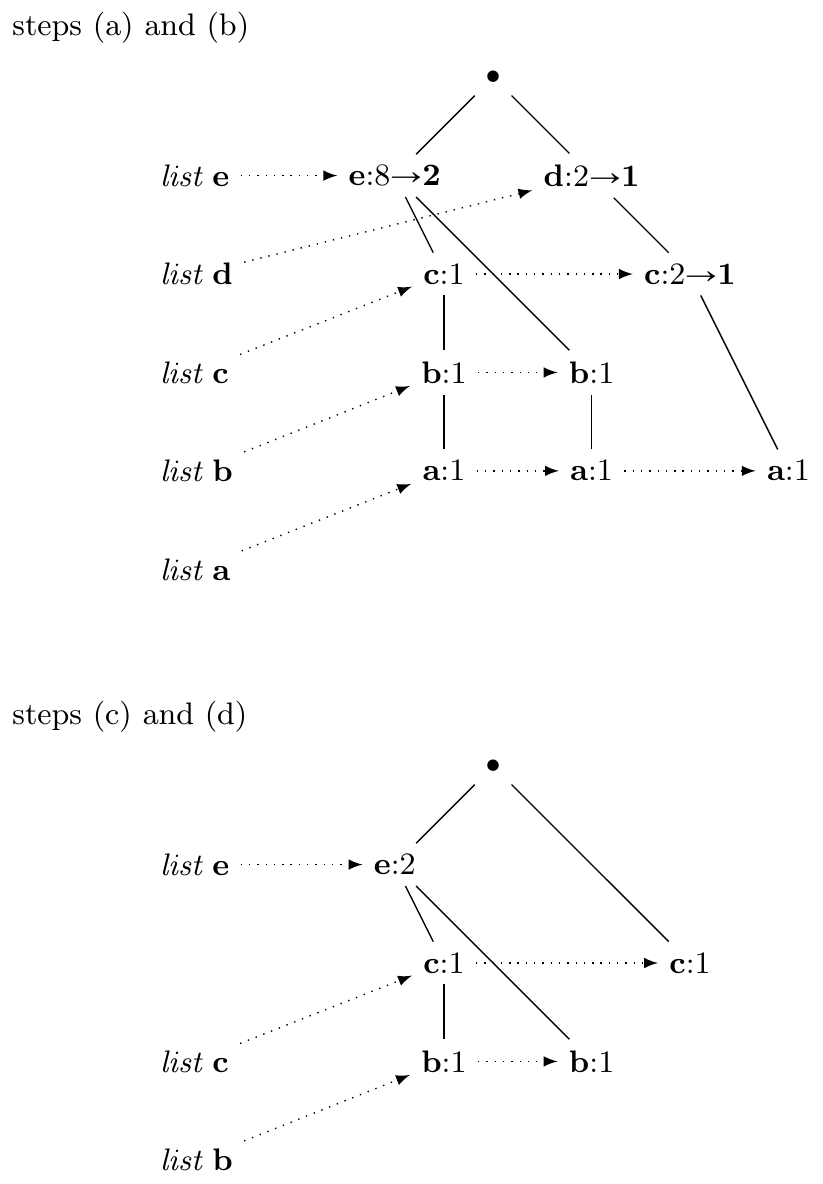}
\end{center}
\caption{Construction of the conditional FP-tree for attribute {\bf a} from the FP-tree in Fig.\,\ref{fig:fp:1}.
\label{fig:fp:conda}}
\end{figure}

\subsection{Complete FP-trees and inner intersections}

Uno {\it et al.} use a so-called {\it complete FP-tree}, where each node contains
a set of attributes on the path from the node up to the root, stored as a bitarray.
See Fig.\,\ref{fig:fp:2} for an example.

\begin{figure}
\begin{center}
\includegraphics{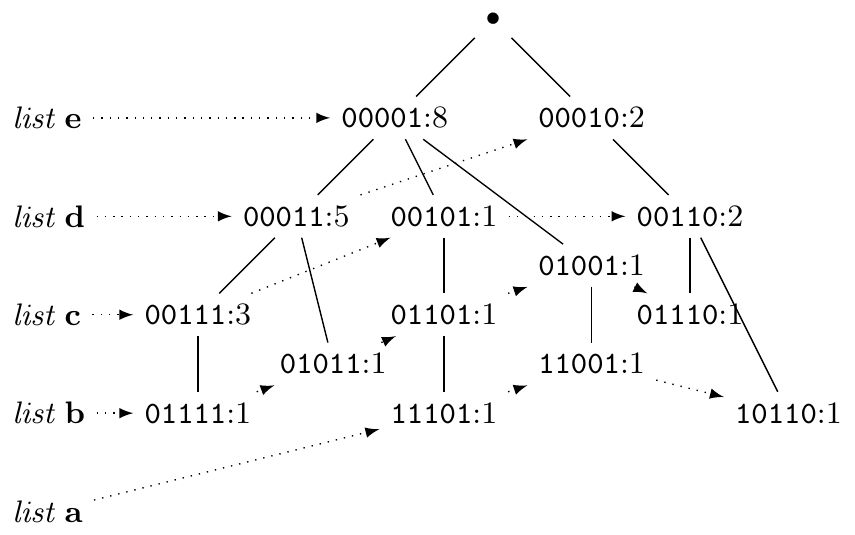}
\end{center}
\caption{Complete FP-tree corresponding to the FP-tree in Fig.\,\ref{fig:fp:1}.\label{fig:fp:2}}
\end{figure}
\noindent
Note that in the complete FP-tree, each node has a unique bitarray, which we can use as an identifier.
Furthermore, for each non-root node, 
we have complete information about the related set. We do need to 
follow the path to the root to find the set.
Therefore, we can omit the tree structure, and store just the lists
(Fig.\,\ref{fig:fp:3}, ignore the rightmost column for now).

\bigskip

The lists play the role of buckets for the occurrence deliver (described in Section~\ref{sec:arraylists}).
For example, the '{\it list}~{\bf c}' in Fig.\,\ref{fig:fp:3} says that the bucket contains $6 =3+2+1$
objects: 
\begin{itemize}
\item
three of them have the attributes {\bf c},{\bf d},{\bf e}; 
\item
two of them have the attributes {\bf c},{\bf d} but not {\bf e}, and
\item
one has the attributes {\bf c},{\bf e} but not {\bf d}.
\end{itemize}
 We have no information about the presence or absence of attributes {\bf a}
and~{\bf b}. We will see later that we actually do not need it here, as we will have 
sufficient 
information in the inner intersection.

\begin{figure}
\begin{center}
\includegraphics{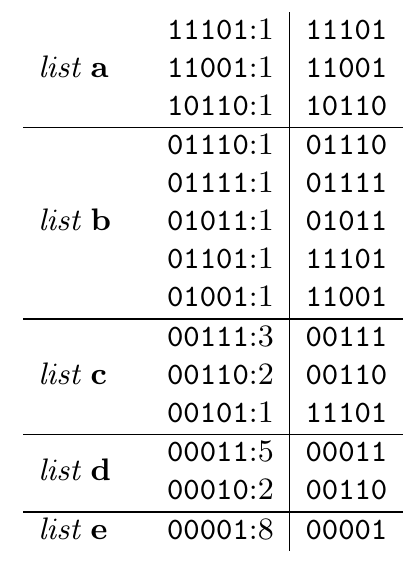}
\end{center}
\caption{Complete FP-tree with inner products.\label{fig:fp:3}}
\end{figure}

While an FP-tree, or complete FP-tree, is sufficient for mining frequent 
itemsets, we need additional information for closed sets. Specifically,
inner intersections (described in Section~\ref{sec:conddat}).
For that reason, we add to each node an intersection (stored as a bitarray) of all rows which it represents (see Fig.\,\ref{fig:fp:3}).

\begin{example}
In our running example, the node \texttt{00111} represents three rows -- namely, 5,6, and 8.
Their intersection is $\{${\bf c},{\bf d},{\bf e}$\}$, therefore, the node has the inner intersection \texttt{00111}.
\end{example}

\subsection{Construction of the complete FP-tree with inner products}

The construction of the complete FP-tree with inner products has two steps.
In the first step (initial step), we put the input data into the corresponding lists.
In the second step (extension step), we extend the items in the list to a path to the root.

\bigskip

\noindent
{\bf Initial step:}
We traverse through the rows of the input data. 
For each row, we find the list which corresponds to its lowest attribute (with respect to the cardinality order).
If there already is a node with the same set in the list, we just increase its counter by one\footnote{or its weight}.
Otherwise, we make a new node for the set; its counter is set to 1 and its inner product 
is the same as the row (an intersection of one set is the same set). The node is put in the list.

\begin{example}
The fifth row in the running example, specifically $\{${\bf c},{\bf d},{\bf e}$\}$ should be put to `{\it list} {\bf c}'.
We create a new node with set (bitarray) \texttt{00111}, its counter being set to 1, and its inner intersection is again \texttt{00111}.
Later, we process the eighth row -- also $\{${\bf c},{\bf d},{\bf e}$\}$. As we already have a node with the set \texttt{00111}
in `{\it list} {\bf c}',
we just increase its counter by one.
\end{example}
Figure~\ref{fig:fp:init} shows an ``unfinished'' FP-tree after the initial step.

\begin{figure}
\begin{center}

\includegraphics{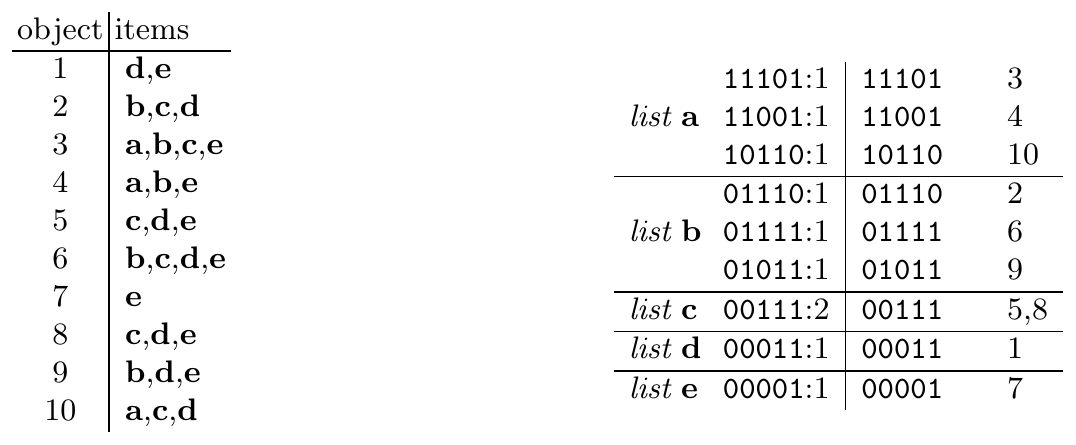}
\end{center}
\caption{``Unfinished'' complete FP-tree with inner intersections after the initial step\label{fig:fp:init}; the numbers on the right side are the numbers of the row in the original data. They are not part of the FP-tree, they are there just for ease of understanding.}
\end{figure}

\bigskip

\noindent
{\bf Extension Step:} For each list in the descending order (with respect to the cardinality order), do the following:
For each node $\langle$set $A$, $k$, $II${}$\rangle$, where 
$A$ is a set, $k$ is a counter, and $II$ is an inner intersection, 
 in the list, make a new set  (bitarray) $A'$ removing the lowest attribute from the set stored in the node.
If the new set $A'$ is non-empty, take the list which corresponds to its lowest new attribute.
Then, if there is a node with the set $A'$ in the list, increase its counter by $k$ and intersect its inner intersection 
with $II$.
Otherwise, make a node $\langle A', k, II\rangle$ and put it in the list.

\begin{example}
\label{ex:extension_step}
Assuming the result of the initial step in Fig.\,\ref{fig:fp:init}, we take sets in the nodes in `{\it list}~{\bf a}'
and remove their lowest attribute (being attribute {\bf a}). We obtain \texttt{01101}, \texttt{01001}, and \texttt{00110}.
The former two belong to `{\it list}~{\bf b}'. 
There are no nodes with these sets in the list, therefore
we create new nodes for them and copy their counters and inner intersection from the original nodes.
The last one belongs to `{\it list}~{\bf c}'. Again, there is no node with the set  \texttt{00110}. Therefore
we put a new node into it.
The intermediate result after extending all nodes in `{\it list}~{\bf a}' is shown in Fig.\,\ref{fig:fp:extend1}\,(left).

Similarly, we continue with nodes in `{\it list}~{\bf b}' (including the new nodes).\
When we process the node with set \texttt{01110}, we remove its lowest attribute and obtain \texttt{00110}.
We see that it belongs to `{\it list} {\bf c}' and there is already a node with this set.
Therefore,
we just increase its counter and intersect its inner intersections with the values from the original node. 
The intermediate result after extending all nodes in `{\it list}~{\bf b}' is shown in Fig.\,\ref{fig:fp:extend1}\,(right).

When this step is finished for all the lists, we obtain the complete FP-tree shown in Fig.\,\ref{fig:fp:3}.
\end{example}

\begin{figure}
\begin{center}
\includegraphics{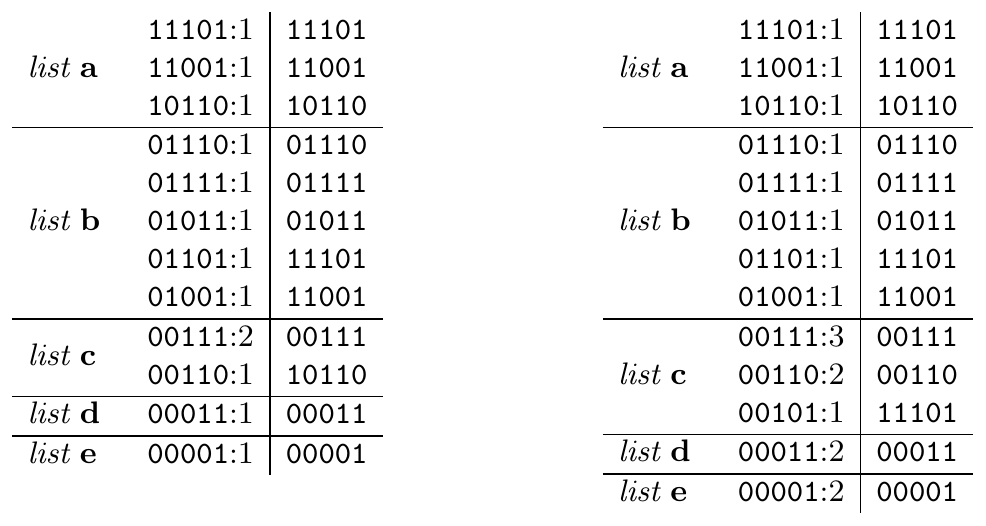}
\end{center}
\caption{
Intermediate result of the construction of the complete FP-tree with inner intersections from Example~\ref{ex:extension_step}
\label{fig:fp:extend1} after extending all nodes in `{\it list} {\bf a}' (left) and `{\it list} {\bf b}' (right).}
\end{figure}

\subsection{Conditional database}

We already said that the lists of a complete FP-tree serve as the buckets for occurrence deliver.
Now, we show that we can easily use them for the construction of the conditional database.

To construct the conditional database for an attribute, simply take its list and perform
the extension step of the construction of the FP-tree, and then remove (or just ignore) the list and the lists of lower attributes. 
If a minimal support is required, we can also remove infrequent lists.

\begin{example}
In our running example, we can extract a conditional database
for attribute {\bf a} by taking the `{\it list} {\bf a}', performing 
the extension step, and removing the `{\it list} {\bf a}'.
The result is shown in Fig.\,\ref{fig:fp:cd}. Note, 
that 
if we remove the infrequent item {\bf d}, 
we receive the same conditional database as in 
the simple FP-tree in Example~\ref{ex:condi}.
\end{example}

\begin{figure}
\begin{center}
\includegraphics{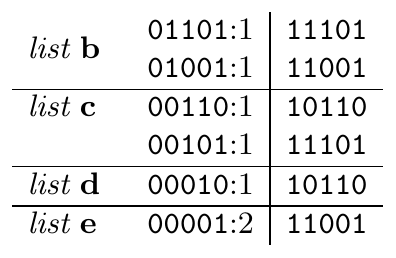}
\end{center}
\caption{Conditional complete FP-tree for attribute {\bf a} with inner intersections extracted from the FP-tree from Fig.\,\ref{fig:fp:3}
\label{fig:fp:cd}}
\end{figure}

\subsection{Reuse of the lists/buckets}
The occurrence deliver (in Section~\ref{sec:arraylists}) 
reused buckets when it used the right to left sweep 
through the search space.
A similar principle can be used with an FP-tree. Indeed, whenever
the algorithm finishes processing a conditional FP-tree, the space allocated
for it can be then used by the next conditional FP-tree.
This is better demonstrated in the next example.

\begin{example}
The conditional FP-tree for {\bf e} is trivial: it is an empty tree.\footnote{This holds generally true; the highest attribute has empty conditional FP-tree.}
When we finish processing the attribute {\bf e}, we can remove its list and use that space for the conditional 
FP-tree of attribute {\bf d}. Note that this conditional FP-tree has only the `{\it list} {\bf e}'.
Figure~\ref{fig:fp:reuse}\,(left) shows the situation where we removed `{\it list} {\bf e}' and used its 
space for the conditional FP-tree of attribute {\bf d}.

Analogously, when we finish processing the attribute {\bf d}, we can delete the `{\it list} {\bf d}' and `{\it list} {\bf e}'
and use their space for the conditional FP-tree of attribute {\bf c}. This is shown in Figure~\ref{fig:fp:reuse}\,(right).

\begin{figure}
\begin{center}
\includegraphics{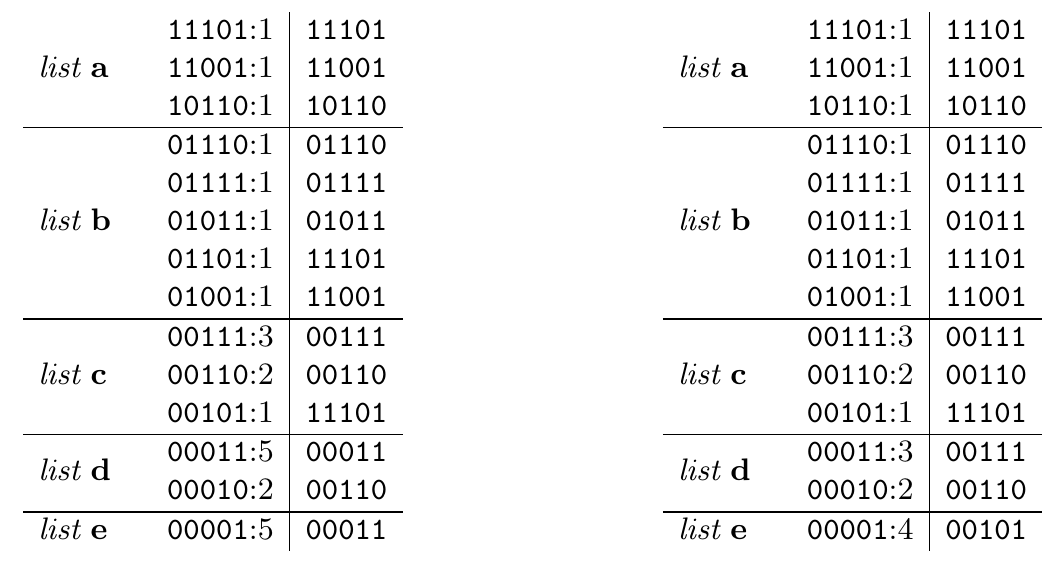}
\end{center}
\caption{Complete FP-tree with inner intersections where 
(left):
we removed `{\it list} {\bf e}' and used its 
space for the conditional FP-tree of attribute {\bf d};
(right):
we removed 
 `{\it list} {\bf d}' and `{\it list} {\bf e}'
and used their space for the conditional FP-tree of attribute {\bf c}.
\label{fig:fp:reuse}}
\end{figure}

\end{example}

\subsection{Extraction of intents}
For each list in a (conditional) complete FP-tree, the intersection of inner intersections of its nodes is an intent. LCM can easily compute the intersections and print them out whenever they satisfy the canonicity test.

\section{Conclusions}
\label{sec:conc}
We analyzed LCM from the point of view of FCA and concluded that it is a CbO-based algorithm
with additional features directed towards processing sparse data. We also compared the additional 
features with those of FCbO and InClose2+ known in the FCA community.

\paragraph{Future research:} We see two main directions for our upcoming research:
\begin{itemize}
\item The investigation of other algorithms for closed frequent itemset mining 
and putting them into context with FCA algorithms.
\item Experimental evaluation of the incorporation of LCM's features in CbO-based algorithms;
this could lead to fast implementations of the algorithms. 
\end{itemize}

\noindent
{\bf Acknowledgment:}
The authors acknowledge support by the grants:

\begin{itemize}
\item
IGA~2020 of 
Palack\'y University Olomouc, No.\,IGA\_PrF\_2020\_019,
\item
JG~2019 of 
Palack\'y University Olomouc, No.\,JG\_2019\_008.
\end{itemize}

%


\end{document}